\documentclass[12pt]{iopart}
\usepackage{iopams}
\usepackage{graphicx}

\usepackage{bm}
\usepackage{multirow}
\usepackage{color}
\usepackage{hyperref}
\usepackage{wrapfig}

\usepackage[sort&compress, numbers, comma]{natbib}

\usepackage[british]{babel}

\begin{document}

\title{Constraints on the SZ Power Spectrum on Degree Angular Scales in WMAP Data} 

\author{Shahab Joudaki}
\address{Center for Cosmology, Department of Physics and Astronomy, 
University of California, Irvine, CA,  92697}
\ead{joudaki@uci.edu}

\author{Joseph Smidt}
\address{Center for Cosmology, Department of Physics and Astronomy, 
University of California, Irvine, CA,  92697}
\ead{jsmidt@uci.edu}

\author{Alexandre Amblard}
\address{Center for Cosmology, Department of Physics and Astronomy, 
University of California, Irvine, CA,  92697}
\ead{amblard@uci.edu}

\author{Asantha Cooray}
\address{Center for Cosmology, Department of Physics and Astronomy, 
University of California, Irvine, CA  92697}
\ead{acooray@uci.edu}

\begin{abstract} 
The Sunyaev-Zel'dovich (SZ) effect has a distinct spectral signature that allows its separation from fluctuations in the cosmic microwave background (CMB) and foregrounds. Using CMB anisotropies measured in Wilkinson Microwave Anisotropy Probe's five-year maps, we constrain the SZ fluctuations at large, degree angular scales corresponding to multipoles in the range from 10 to 400. We provide upper bounds on SZ fluctuations at multipoles greater than 50, and find evidence for a hemispherically asymmetric signal at ten degrees angular scales. The amplitude of the detected signal cannot be easily explained with the allowed number density and temperature of electrons in the Galactic halo. We have failed to explain the excess signal as a residual from known Galactic foregrounds or instrumental uncertainties such as $1/f$-noise. 
\end{abstract}

\maketitle

\section{Introduction}

The Sunyaev-Zel'dovich (SZ) effect~\cite{Sunyaev:1972eq} is
the inverse Compton scattering of the cosmic microwave background (CMB)
by electrons throughout the universe. The  SZ effect can be partitioned into two components: the 
kinetic effect (also known as the Ostriker-Vishniac effect~\cite{Ostriker86}) due to bulk motion of electrons with 
respect to the rest frame of the microwave background, and the thermal effect due 
to energy transfer from hot electrons in massive galaxy clusters~\cite{Komatsu:1999ev, Cooray:2000ge, Molnar:2000de, Springel01, Seljak:2001rc, Sadeh:2004jk}. The SZ thermal signal is expected to be important in future constraints on the underlying cosmology of the universe, by  revealing the properties of clusters in cluster-counting experiments (e.g. see Carlstrom et al. 2002~\cite{Carlstrom:2002na}) and via the angular power spectrum~\cite{Komatsu:2002wc}.

At degree angular scales, there are no detections of the SZ fluctuations as primordial anisotropies of the CMB dominate the SZ signal by at least 3 orders of magnitude. Fortunately, a separation of the SZ anisotropies from those of the CMB is possible due to the fact that the SZ effect has a  distinct  frequency spectrum, as the inverse-Compton scattering on average increases the net energy of the CMB photons and move photons from the low frequency Rayleigh-Jeans (RJ) tail to higher frequencies~\cite{Sunyaev:1972eq}. In multi-frequency CMB experiments spanning a wide range of frequency coverage across the SZ null frequency at 217 GHz, one can separate the CMB down to sub-percent level required to study the SZ anisotropy power spectrum~\cite{Cooray:2000xh}. This approach was first applied to constrain the sub-degree SZ fluctuations in BOOMERanG 2003 data~\cite{Veneziani:2009es}. Here, we analyze the WMAP five-year CMB data to constrain SZ anisotropies at degree angular scales corresponding to multipoles $\ell=10-400$ on the sky. We ignore $\ell < 10$ as a measure to avoid biases from previously reported anomalies, the divergence of 1/$f$-noise, and large uncertainties associated with simulating the CMB sky properly to match WMAP data at these multipoles~\cite{ben92,hin96,copi04,land05,schwarz04,oliveira04,eriksen04,bennett10}.

In Section 2 we detail our approach to extract the SZ signal from the multifrequency 
WMAP5 data set; in Section 3 we present our results; and in Section 4 we provide a discussion of these results.

\section{CMB, Noise, and Foreground Removal}
\label{sec:analysis}

We distinguish the SZ signal from other sources of anisotropies by minimizing the covariance relative to the SZ frequency dependence (also see Veneziani et al 2009~\cite{Veneziani:2009es}). Although this technique is originally employed for removing foregrounds from CMB anisotropies~\cite{Tegmark:1995pn, Tegmark:2003ve, Amblard:2006ef}, we utilize it to recover SZ fluctuations from which the primordial CMB is subtracted along with other sources of noise.

We measure the raw SZ power spectrum from a weighted mean of the spectra in different frequency bands~\cite{Cooray:2000xh}:
\begin{equation}
\label{eq:csz}
C_\ell^{\rm{raw SZ}} = \textbf{w}_{\ell}^T\textbf{C}_{\ell}\textbf{w}_{\ell},
\end{equation}
where the scale dependent weights $\textbf{w}_\ell$ at each frequency $\nu$
are obtained by minimizing the contribution of the CMB, noise, and foregrounds in the variance ${C}_{\ell}^{\rm{raw SZ}}$ (of multipole moments $a_{{\ell}m}^{\rm{raw SZ}} = \sum_i {{w_\ell(\nu_i)} \over {s(\nu_i)}} a_{{\ell}m}(\nu_i)$ as elucidated below). 
We enforce the constraint that SZ estimation is unbiased, via $\sum_i w_{\ell}(\nu_i) = 1$. 
The optimal weights for reconstructing the SZ are thereby obtained from applying the method of Lagrange multipliers to Eqn.~\ref{eq:csz}:
\begin{equation}
\label{eq:weight}
\textbf{w}_{\ell} = \frac{\textbf{C}_{\ell}^{-1}\textbf{e}}{\textbf{e}^T\textbf{C}_{\ell}^{-1}\textbf{e}},
\end{equation}
where $\textbf{e}$ is a unit vector, such that $e({\nu_i}) = 1~\forall \nu_i$.

We construct the positive-definite covariance matrix for frequency bands $i$ and $j$ as a function of angular scale, wherein the SZ frequency dependence is removed to prevent its minimization:
\begin{equation}
\label{eq:covar}
\textbf{C}_{(ij)}(\ell) = \frac{{\langle a_{\ell m}^i a_{\ell m}^{j*} \rangle}/{f_{\rm{sky}}}}{s(\nu_i)s(\nu_j)},
\end{equation}
where $f_{\rm sky}$ is the sky fraction analyzed, the multipole moments $a_{\ell m}$ are extracted from either data or simulated maps at separate frequencies, and $s(\nu) = 2 - (x/2)\rm{coth}(x/2)$, with $x \equiv h\nu/kT_{\rm{CMB}} \approx \nu/56.8~\rm{GHz}$, is the frequency dependence of the SZ at the center of each of the WMAP frequency bands relative to the CMB. 
In the RJ limit $s(\nu) \rightarrow 1$, such that $C_{\ell}^{\rm{SZ}}(\nu,\nu') = s(\nu)s(\nu')C_{\ell}^{\rm{SZ}}$, where $C_{\ell}^{\rm{SZ}}$ is the SZ power spectrum in the RJ limit.  For the two preferred WMAP5 bands $[{\rm V}, {\rm W}] = [60.8, 93.5]$ GHz, we find
$s(\nu) = [0.906, 0.784]$. For the data, the multipole moments are obtained from WMAP five-year maps 
\begin{equation}
a_{{\ell}m;\nu}^{\rm{data}} = a_{{\ell}m;\nu}^{\rm WMAP}/b_{\ell;\nu},
\label{almeqndata}
\end{equation}
normalized by the measured beam window function $b_{\ell;\nu}$ in the respective frequency band.

To compute the covariance matrix of non-SZ sources, we Monte-Carlo simulated a set of 250 Gaussian sky maps with HEALPix\footnote{http://healpix.jpl.nasa.gov} based on the CAMB-generated~\cite{Lewis:1999bs} input power spectrum in accordance with the WMAP5 best-fit $\Lambda$CDM cosmology ($[\Omega_ch^2, \Omega_bh^2, \Omega_\Lambda, n_s, \tau, \Delta_R^2] = [0.1099, 0.02273, 0.742, 0.963, 0.087, 2.41 \times 10^{-9}]$). We find consistent results with simulations using the exact WMAP5 measured CMB power spectrum in each of the bands. From these frequency independent Gaussian maps, masked along the Galactic plane via KQ75\footnote{http://lambda.gsfc.nasa.gov/product/map/dr3/} (admitting 71.6\% of the sky), we then extracted the moments $a_{lm}^G$ up to $\ell = 400$ with HEALPix. To account for the masking in our estimated spectra, we used the algorithm described by Hivon et al (2002)~\cite{Hivon02}.

\begin{figure}[!t]
\vspace{-0.5em}
\begin{center}
\centerline{\includegraphics[scale=0.70]{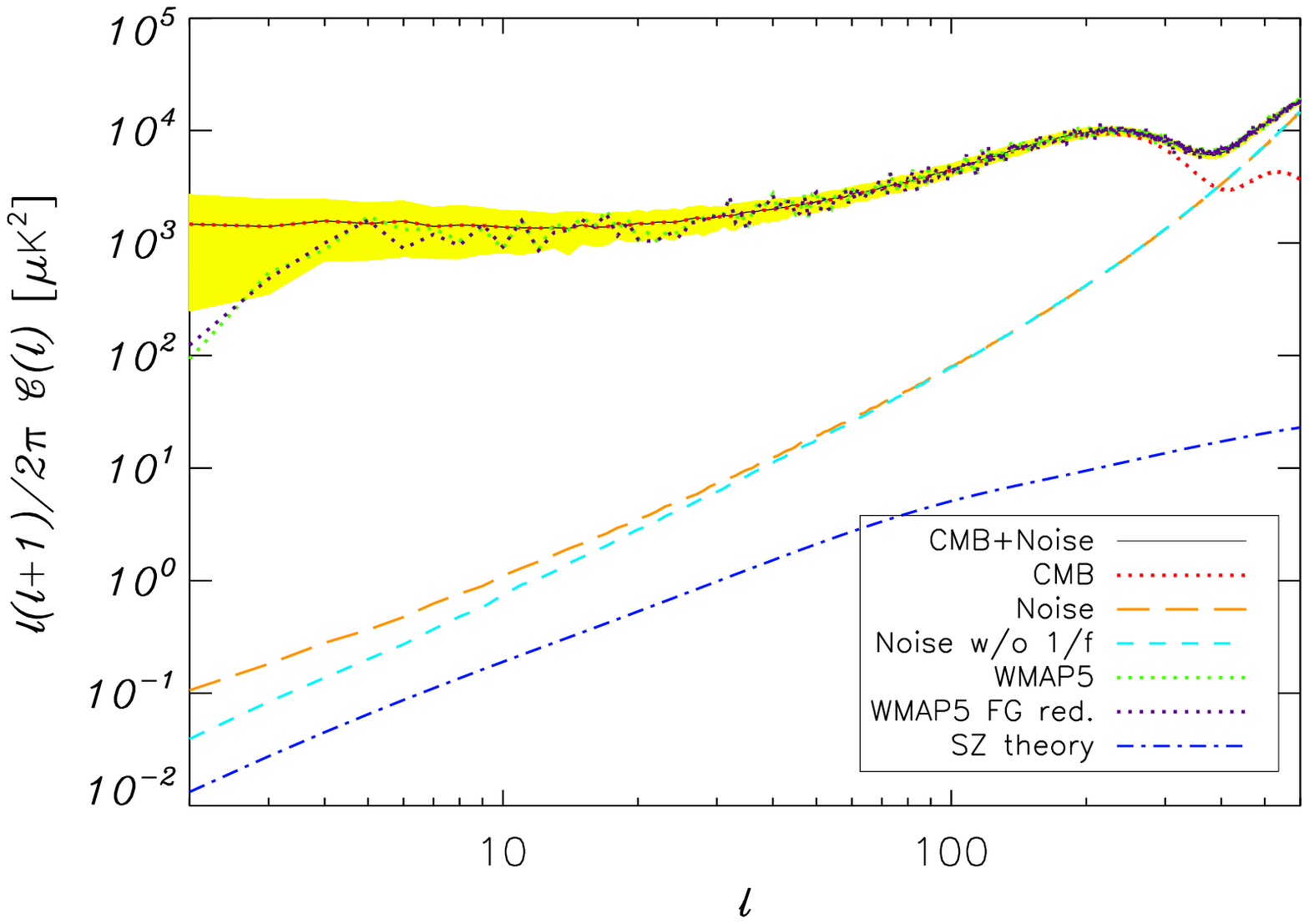}}
\vspace{-2.0em}
\centerline{\includegraphics[scale=0.70]{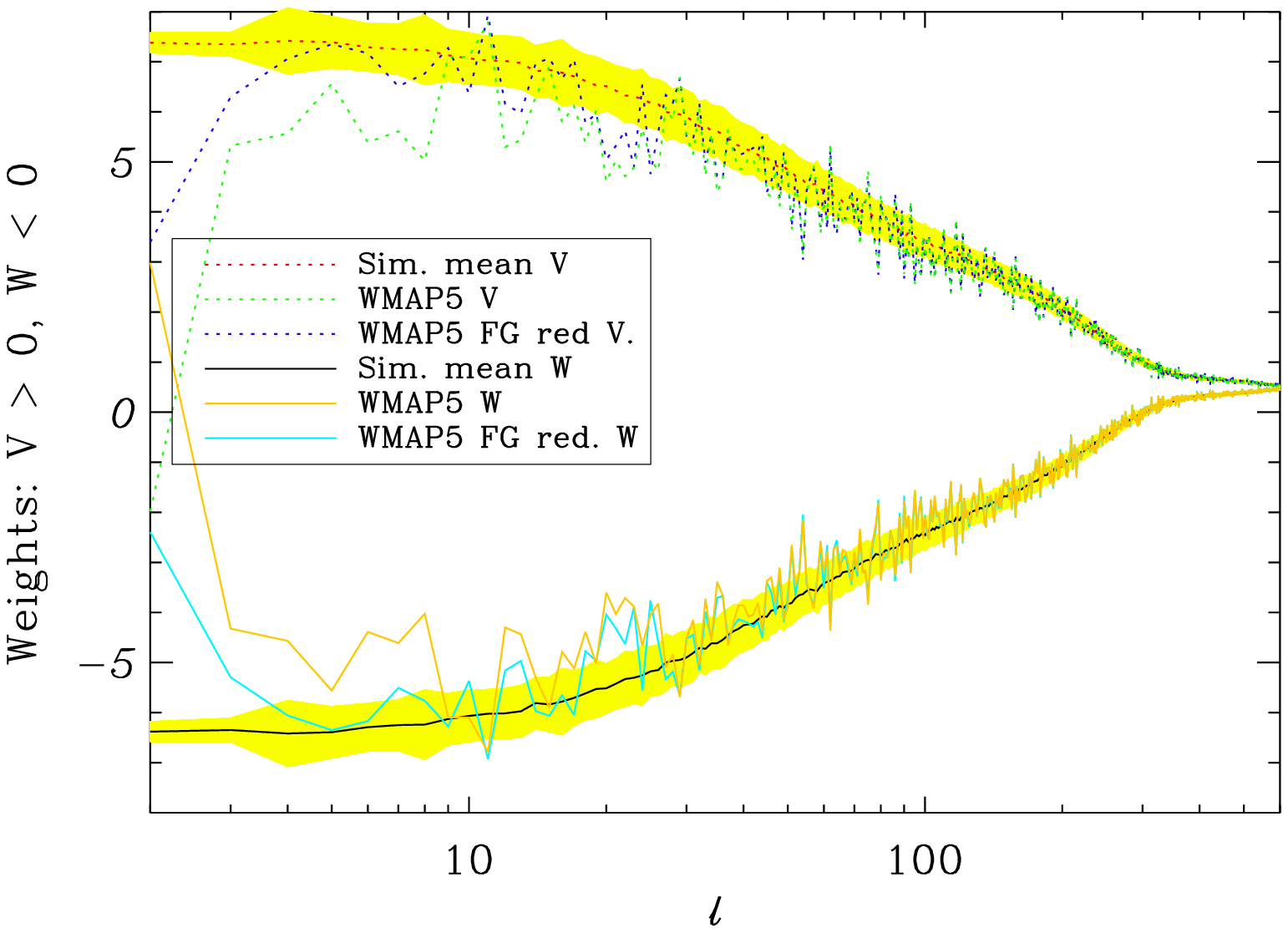}}
\vspace{-1.0em}
\caption{{\it Top}:~The average simulated auto spectrum (Eqn.~\ref{eq:covar}) in the W frequency band for the combination of Gaussian CMB and noise illustrated by a black solid line about its 1$\sigma$ yellow error band. The red dotted line stems from pure CMB, and the orange long-dashed line is the noise contribution (beam normalized), while the dashed cyan line solely encapsulates the white noise. The purple and green dotted lines are the spectra obtained from the foreground reduced and unreduced WMAP5 maps, respectively. The dot-dashed blue line represents the signal from the theoretical Komatsu-Seljak SZ power spectrum~\cite{Komatsu:2002wc}. 
{\it Bottom}:~Optimal weights constructed from the covariance matrix following Eqn.~\ref{eq:weight}. The upper set of dotted lines are obtained from the V-band and the lower set of solid lines from the W-band. The mean simulated weights run through the yellow bands, representing the 1$\sigma$ uncertainty about the mean. The foreground cleaned and uncleaned weights from WMAP5 are also illustrated for each frequency band.}
\label{fig:szcovarandweight}
\end{center}
\vspace{-1.0em}
\end{figure}

In addition to these CMB simulations, we create 250 noise maps for each frequency band, given by
\begin{equation}
\label{eq:noise}
N_{\nu}(\hat{\textbf{n}}) = {\sigma_{0; \nu} \over {\sqrt{N_{{\rm obs};\nu}}}} n(\hat{{\textbf n}}),
\end{equation} 
where $N_{\nu}(\hat{\textbf{n}})$ is the noise map, $n(\hat{{\textbf n}})$ is a white noise map, the frequency dependent $N_{\rm obs}$ is the number of observations per pixel, and the noise per observation $\sigma_0 = [3.133, 6.538]$ mK for the $[\rm{V, W}]$ bands\footnotemark[\value{footnote}]. We moreover explore the impact of $1/f$-noise in our simulations. To this end, we assume similar 1/$f$-noise properties for five-year data as for one-year data consistent with the observations (see Eqn.~8 and Table~1 of Hinshaw et al. 2003~\cite{Hinshaw:2003ex}), and take the same amplitude ratio between one-year data and five-year data of 1/$f$-noise as the corresponding amplitude ratio of the white noise~\cite{ekom}.
On the largest angular scales, the $1/f$-noise boost in the V-band is a factor of two, and in the W-band the boost is at most a factor of three (see Fig.~\ref{fig:szcovarandweight}) relative to the white noise amplitude. We comment on our result with and without $1/f$-noise included in the discussion.
From the set of noise-generated maps, we produce the corresponding multipole moments, $a_{{\ell}m;\nu}^{\rm N}$, out to the same angular scale as their CMB counterparts. We correct for the beam according to $a_{{\ell}m;\nu} = a_{{\ell}m}^{\rm{G}} + a_{{\ell}m;\nu}^{\rm N}/b_{\ell;\nu}$.

The WMAP5 maps represent a weighted mean of the five single-year maps~\cite{Gold:2008kp}. In subtracting from the unreduced maps synchrotron and free-free emission, primarily at frequencies below 60 GHz, and thermal dust emission which dominates at frequencies above 60 GHz, the WMAP team~\cite{Hinshaw:2006ia, Page:2006hz, Gold:2008kp} utilized a temperature difference map of the K (22.8 GHz) and Ka (33.0 GHz) channels, an extinction-corrected H$\alpha$ map of Finkbeiner (2003)~\cite{Finkbeiner:2003yt}, and dust emission Model 8 of Finkbeiner et al (1999)~\cite{Finkbeiner:1999aq}.

We account for potential systematic uncertainties in the WMAP team's foreground removal by adding to the multipole moments of the CMB and noise simulations a contribution from foreground residuals. We further consider 1\% uncertainties in the beam transfer functions of each frequency band~\cite{Hill:2008hx}. Consequently, the multipole moments from simulations take on the final combined form
\begin{equation}
a_{{\ell}m;\nu}^{\rm{sim}} = a_{{\ell}m}^{\rm G} + a_{{\ell}m;\nu}^{\rm N}/{B_\nu(\ell)} + a_{{\ell}m}^{\rm d}D_\nu + a_{{\ell}m}^{\rm ff}F_\nu + a_{{\ell}m}^{\rm sync}S_\nu ,
\label{almeqn}
\end{equation}
where $B_\nu(\ell) = b_{\ell;\nu} \left({1+\Delta{b} \gamma^b_\nu(\ell)}\right)$, $D_\nu = d_\nu (1+\Delta{d_\nu} \gamma^d_\nu)$, $F_\nu = f_\nu (1+\Delta{f_\nu} \gamma^f_\nu)$, and $S_\nu = s_\nu (1+\Delta{s_\nu} \gamma^s_\nu)$, such that $\gamma$ is a frequency dependent, but multipole independent unless otherwise specified, Gaussian random number drawn for each of the 250 simulations. The frequency dependent $d_\nu$, $f_\nu$, and $s_\nu$ are the WMAP foreground template amplitudes of the considered dust, free-free, and synchotron emissions, respectively. These amplitudes are given in Table~2 of Gold et al (2009)~\cite{Gold:2008kp}. The dimensionless $\Delta{b}$, $\Delta{d_\nu}$, $\Delta{f_\nu}$, and $\Delta{s_\nu}$ respectively encapsulate a 1\% beam uncertainty, and a [4.4, 5.6]\% uncertainty on the amplitudes of the dust, free-free, and synchrotron foreground templates in the [V, W] bands. 

In practice, we only simulate the deviation from the WMAP foregrounds and utilize the foreground cleaned maps in our analysis. The synchrotron foreground template was constructed by the difference of the WMAP5 temperature map in the K band with that of the Ka band, whereas the other two templates were obtained from the LAMBDA\footnotemark[\value{footnote}] website. The $\sim5\%$ foreground template uncertainties were obtained from the stated 15~${\mu}$K uncertainty on the removal of the combined foregrounds~\cite{Gold:2008kp}. More specifically, we add all of the foreground templates together (multiplied by the respective amplitudes in each band), calculate the dispersion $\sigma_t$ of the combined template pixels, and let the fractional error on each foreground amplitude equal ${{15~{\mu{\rm K}}} \over {\sigma_t}}$. Table~\ref{tableone} shows the foregrounds render a bias and increase the uncertainty in the SZ estimation primarily on large scales ($\ell < 50$), while the beam uncertainty is manifested on the smallest scales ($\ell > 150$).

In addition to our account of residual foregrounds in the error analysis, we add a flat spectrum of radio point sources to the temperature power spectrum of the above simulations, such that the numerator of Eqn.~\ref{eq:covar} becomes $C_{\ell}^{\rm sim} \rightarrow C_{\ell}^{\rm{G+N+FG_{res}}} + C^{\rm{ps}}$, via the frequency dependent model for the point source power spectrum~\cite{Nolta:2008ih}:
\begin{equation}
\label{eq:psource}
C^{\rm ps} (\nu_i, \nu_j) = A_{\rm ps}{{r(\nu_i) r(\nu_j)}\over{b(\nu_i)b(\nu_j)}} \left ({{\nu_i \nu_j} \over {\nu_Q^2}} \right )^{\alpha-2},
\end{equation} 
where $\nu_i$ and $\nu_j$ are the central frequencies of the two considered bands, normalized by that of the Q band $\nu_Q = 40.7$ GHz. The point source amplitude $A_{\rm{ps}} = (11.1 \pm 4.1) \times 10^{-3}~\rm{\mu}\rm{K}^2\rm{sr}$ for the band combination VW, and the spectral index of the point source flux $\alpha = -0.09 \pm 0.18$~\cite{Nolta:2008ih}. Moreover,
\begin{equation}
\label{eq:psr}
r(\nu) = {{(e^x - 1)^2} \over {x^2e^x}} ,
\end{equation} 
where as before $x \equiv h\nu/kT_{\rm{CMB}}$. Beyond point source removal, we account for the uncertainty in the amplitude and spectral index by independent Gaussian realizations around the mean. In Table~\ref{tableone}, we see that point sources primarily bias our results on small scales. We have ignored clustering of point sources, which may add some further power on these small scales.

The multipole ranges for our bins are listed in Table~\ref{tableone}. We obtain the binned spectra by weighting each $C_\ell$ (averaged over the 250 simulations for the simulated case) with the respective uncertainty $\sigma_\ell$ at that multipole.

\section{Results}
\label{sec:results}

We show the contributions to the power spectrum from the CMB, noise, and their combination, along with the WMAP five-year spectrum and the theoretical Komatsu-Seljak SZ power spectrum~\cite{Komatsu:2002wc} for the W band in Fig.~\ref{fig:szcovarandweight} (top). The WMAP data trace the simulated spectra closely, which become shot noise dominated for $\ell > 350$. 

In Table~\ref{tableone}, we state the binned weights of the fore- ground reduced WMAP5 data for the two frequency bands.
The optimal weights at each multipole are illustrated in Fig.~\ref{fig:szcovarandweight} (bottom) for both the data and simulations. The two bands have weights roughly symmetrically aligned along zero, and decrease towards smaller angular scales. The residuals are the average spectra measured on our SZ-free simulations and represent our bias. The total residual is slightly different from the sum of the partial residuals due to correlation between components. The uncertainties in the table are the $2\sigma$ dispersion $\sigma_{\rm bin}$ measured with our simulations. The final SZ spectrum values are corrected from the residual bias. The total error budget is given by the rms of the partial uncertainties. While in the second and third bins we find no detectable signal, in the first bin with $10 < \ell <50$, we find an excess signal of $58.6 \pm 17.3$ $\mu$K$^2$ (95\% c.l.). 

\begin{figure}[!t]
\centerline{\includegraphics[scale=0.70]{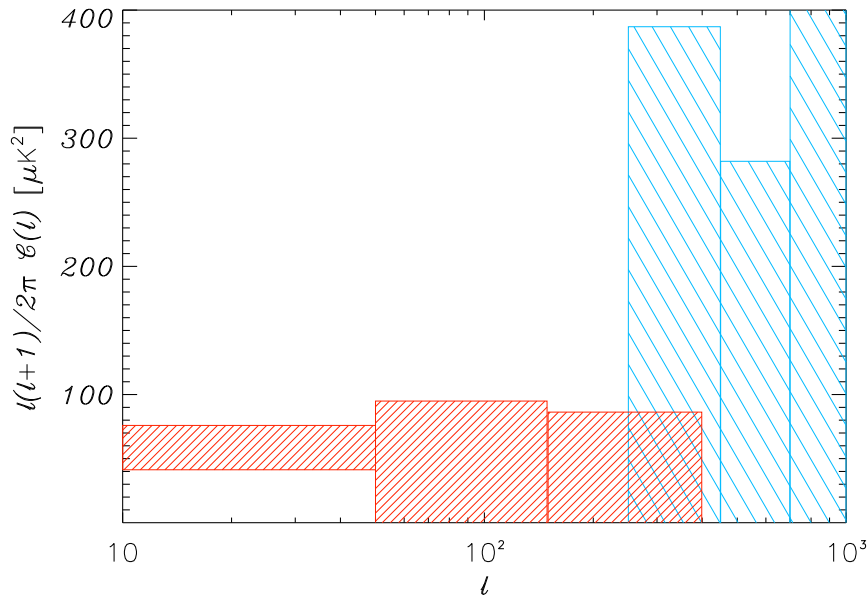}}
\vspace{-1.0em}
\caption{95\%-level constraints on the SZ spectrum from WMAP five-year data in the [V, W] frequency bands (thinly-sliced red bars), along with smaller-scale BOOMERanG 2003 data in thickly-sliced blue bars from Veneziani et al (2009)~\cite{Veneziani:2009es}.}
\label{fig:szjanmany}
\end{figure}

\begin{table}[!t]
\begin{center}
{\sc SZ Power Spectrum Estimates}
\begin{tabular}{lccc}
\hline
& Bin 1 & Bin 2 & Bin 3\\
$\ell$-range & 10 -- 50 & 50 -- 150 & 150 -- 400\\
\hline \hline
\multicolumn{4}{c}{Optimal weights}\\
\hline
$w_{\rm 60.8 GHz}$ & 5.95 & 3.69 & 1.08\\
$w_{\rm 93.5 GHz}$ & -4.95 & -2.69 & -0.0831\\
\hline \hline
Raw SZ & 259 & 1740 & 4850\\
\hline \hline
\multicolumn{4}{c}{Residuals}\\
\hline
CMB \& Noise & 139 & 1650 & 4690\\ 
Foregrounds & 57.4 & 26.7 & 35.5\\ 
Beam & -0.0414 & -0.754 & 7.97\\
Point sources & 3.08 & 21.9 & 77.5\\
\hline
Total residual & 201 & 1700 & 4810\\ 
\hline \hline
\multicolumn{4}{c}{SZ Band Power Uncertainties}\\
\hline
CMB \& Noise & 15.5 & 50.3 & 49.7\\ 
Foregrounds & 7.47 & 1.82 & 0.948\\
Beam & 0.912 & 2.00 & 4.63\\
Point sources & 0.933 & 3.75 & 5.85\\
\hline\hline
Final SZ Power&58.6$\pm$17.3&44.4$\pm$50.5&36.0$\pm$50.3\\
\hline
\end{tabular}
\end{center}
\caption{The table derives the SZ angular power spectrum $\ell(\ell+1)C_{\ell}/2\pi$ at the RJ end of the frequency spectrum. All values are tabulated in units of $\mu$K$^2$, except for the unitless weights. ``Raw'' and ``Final'' SZ refer to the spectrum pre and post subtraction of the residuals. The Raw SZ spectrum is measured from data, whereas we simulate the contribution from the CMB, instrumental noise, point sources, foreground subtraction uncertainties, and beam uncertainties. The residual foregrounds row quantifies the residual from the sum of the WMAP estimated foregrounds and the deviation from their foreground estimate. These results do not include $1/f$-noise, see the text for discussion. The uncertainties are given at 95\% c.l.}
\label{tableone}
\end{table}

For comparison, if we include a $1/f$-noise contribution in our simulations by a large-scale boost to the white noise (by a factor of 2 to 3), the first bin decreases to $40.3 \pm 20.1$ $\mu$K$^2$ (95\% c.l.). Removing the channels most susceptible to $1/f$-noise (W2 and W4 via Hinshaw et al. 2003~\cite{Hinshaw:2003ex}) does not statistically modify the signal from that originating from the analysis without $1/f$-noise included. Unless $1/f$-noise is underestimated by an order of magnitude, it seems unlikely that excess noise can explain away the detected excess signal. Our findings are robust to an analysis of single-year maps in separate differencing assemblies (whereby Eqn.~\ref{eq:covar} is extended to an average over all combinations of differencing assemblies in bands $i$ and $j$ for all auto and cross-years), whereas a cross-correlation only analysis leads to a non positive-definite covariance.

In Fig.~\ref{fig:szjanmany} we show our final estimate of the SZ-like power spectrum in distinct multipole bins for a combination of the [V, W] bands. There seems to be a clear non-zero signal at the largest angular scales ($10 <\ell <50$). Although foregrounds have already been reduced, this signal could potentially be due to further systematic uncertainties in the foreground removal. The limits on the SZ signal from a 2003 BOOMERanG flight~\cite{Masi06, Veneziani:2009es} have been included for comparison. 
In addition, if we modify Eqn.~\ref{eq:csz} from $C_\ell^{\rm{raw SZ}} = \textbf{w}_{\ell}^T\textbf{C}_{\ell}\textbf{w}_{\ell}$ to include a noise-subtraction, such that $C_\ell^{\rm{raw SZ}} = \textbf{w}_{\ell}^T\left({\textbf{C}_{\ell} - \langle{\textbf{N}_{\ell}}\rangle}\right)\textbf{w}_{\ell}$, then our result in the first multipole bin is altered from $58.6 \pm 17.3$ $\mu$K$^2$ to a value of $69.1 \pm 18.5$ $\mu$K$^2$ (95\% c.l.).

As a simple consistency check on our calculation, we computed the leftover signal from the difference of the V and W maps, i.e. the power spectrum of a map given by ${{\rm{V}(\hat{\textbf{n}})-\rm{W}(\hat{\textbf{n}})} \over {s_V-s_W}}$ and normalized by the beam (the beam is the same to sub-percent level for the two frequency bands at $\ell<50$), rendering a value of 697 $\mu$K$^2$ in the first bin. We subtracted from this quantity the mean simulated noise on the difference map, given by ${{\left\langle (\sigma_{V; \ell}/b_{V; \ell})^2 \right\rangle + \left\langle (\sigma_{W; \ell}/b_{W; \ell})^2 \right\rangle} \over {(s_V-s_W)^2}} = 536~\mu\rm{K}^2$. The signal derived from this method is then $161 \pm 49$ (95\% c.l.) in the first bin. This derivation, however, is sub-optimal since it does not fully minimize frequency dependent quantities such as instrumental noise and foregrounds, but only removes common signals like the CMB. Simulations maintain that this method leads to a higher ``SZ'' signal. Nevertheless, this suggests that the SZ-like signal we have detected with our optimal method is not associated with a residual CMB that we have not accounted for in simulations related to Table~\ref{tableone}.

In binning the SZ spectra into a single multipole bin, we obtain $\ell(\ell+1)C_{\ell}/2\pi = 55.1 \pm 15.6$ $\mu$K$^2$ (95\% c.l.) for the central multipole of $\ell = 205$, which is heavily dominated by the first bin due to its comparatively much tighter error bar. This implies a component with 3\% the value of the primordial CMB in the power spectrum (or 17\% in $a_{{\ell}m}$ temperature). We have failed to find a combination of foreground residuals, beam errors, calibration uncertainty, and similar instrumental systematics that will add up to the needed correction. 

As a test on the reliability of the signal detected in the first multipole bin, we varied the foreground model involving the amplitudes of the synchrotron, free-free, and dust, but within the uncertainties allowed by WMAP data we failed to find a model that produced a null SZ signal in the first bin. However, a null signal can be obtained if the WMAP estimated foreground emission is increased by a factor of two.
Our error budget includes the full uncertainty in the foregrounds to the extent that foregrounds have been established with WMAP data~\cite{Gold:2008kp}. It could be that our signal is a signature of an unknown foreground. If that were to be the case, CMB observations that span a wider range of frequencies will become necessary, and we motivate a study with the ongoing Planck experiment.

\begin{figure}[!t]
\vspace{-0.5em}
\centerline{{\includegraphics[scale=0.54]{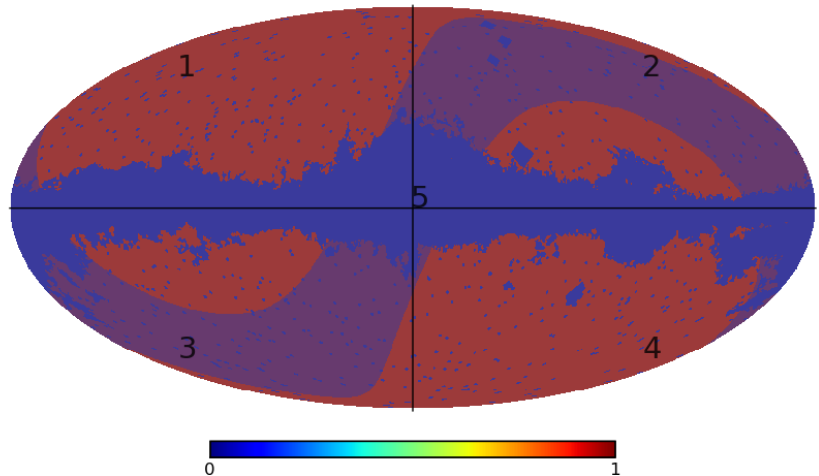}}}
\caption{CMB mask in Galactic coordinates demonstrating the localization choices. The four quarters of the map are numbered accordingly, and the fifth cut is defined by the transparent blue band along the Ecliptic plane ($\pm 20^{\circ}$). The Galactic plane is always masked out with KQ75 in our analysis.}
\label{fig:cmbplot}
\end{figure}

\begin{table}[!t]
\begin{center}
{\sc Large-Scale Signal Localization}
\begin{tabular}{cc}
\hline
Bin 1 ($\ell$ = 10 -- 50) & Final SZ Power\\
\hline\hline
R1 ($f_{\rm sky}$ = 0.177) & 14.9$\pm$32.4\\
R2 ($f_{\rm sky}$ = 0.185) & 49.2$\pm$42.8\\
R3 ($f_{\rm sky}$ = 0.181) & 86.8$\pm$45.3\\
R4 ($f_{\rm sky}$ = 0.173) & 58.9$\pm$30.8\\
R5 ($f_{\rm sky}$ = 0.245) & 49.2$\pm$40.3\\
R1+R2 ($f_{\rm sky}$ = 0.361) & 34.6$\pm$24.8\\
R3+R4 ($f_{\rm sky}$ = 0.355) & 84.9$\pm$24.2\\
R1+R2+R3+R4 ($f_{\rm sky}$ = 0.716) & 58.6$\pm$17.3\\
R1+R2+R3+R4-R5 ($f_{\rm sky}$ = 0.471) & 57.6$\pm$19.6\\
\hline
\end{tabular}
\end{center}
\caption{R1 denotes the north-western quarter of Fig.~\ref{fig:cmbplot} in Galactic coordinates, R2 denotes the north-east, R3 the south-west, R4 the south-east, and R5 the Ecliptic plane in a $\pm 20^{\circ}$ band. The uncertainties are given at 95\% c.l.}
\label{tabletwo}
\end{table}

In order to better understand the physical origin of the measured signal, we repeated our analysis for a number of sky cuts. In Fig.~\ref{fig:cmbplot} we show the CMB mask in Galactic coordinates demonstrating the localization choices. Table~\ref{tabletwo} determines the measured signal to primarily lie in the southern hemisphere, at $84.9 \pm 24.2~\mu\rm{K}^2$ (95\% c.l.). The statistical significance of the signal is equivalent in the two quarters of the southern sky, whereas the north-western quarter is completely devoid of an SZ signal. The excess of power in the southern hemisphere (Ecliptically aligned) was first detected by Eriksen et al (2003)~\cite{Eriksen:2003db} and hypothesized to be due to thermal SZ for $\ell < 10$ by Abramo et al (2006)~\cite{Abramo:2006hs}. Degree scale anomalies in the CMB have previously been found along the Ecliptic plane in Diego et al (2010)~\cite{Diego:2009wi}, whereas we find the signal in that region present at a 2.4$\sigma$-level.

\section{Discussion}
\label{sec:discussion}

If the excess signal we have detected at ten degrees angular scales is indeed SZ, it is clear that the signal does not originate from galaxy clusters at high redshift. Such a signal would lead to a power spectrum that increases with multipole, rendering an SZ signal in the smaller scale bins. A large SZ signal associated with the Extragalactic sky is also ruled out by existing arcminute-scale CMB experiments. We do not see the signal coming from unresolved point sources or Zodiacal light emission as they do not amount to more than a few $\mu$K in strength~\cite{Diego:2009wi}. 

One possibility is that the signal is associated with sub-keV gas in the Galactic halo, explaining why the signal is primarily at small multipoles. 
We can try to address if such a signal is possible based on requirements on the number density of electrons and the existing spectral distortion limits from FIRAS. Since the SZ effect is $-2y$ at RJ frequencies, where $y$ is the Compton $y$-parameter, we can convert the amplitude of the fluctuation power, for a Galactic-like signal with power-law $C_{\ell} \simeq A \ell^{-3}$~\cite{Lagache2007} and assuming a dispersion in $y$ equal to its average value, to obtain $y = (5.1 \pm 0.8) \times 10^{-6}$ (95\% c.l.). This is consistent with the 95\%-level FIRAS constraint of $|y| \leq 1.5 \times 10^{-5}$ from the COBE satellite~\cite{Fixsen:1996nj}.

The electron column density required to render this $y$-parameter in our Milky Way halo can be obtained from 
$N_e \simeq {{y m_ec^2} \over {\sigma_T k_{\rm B}T_e}}$, where $m_e$ is the electron mass, $T_e$ is the electron temperature, with other constants having the usual definition. For a 0.3 keV electron gas temperature, we find $N_e = {1.3 \pm 0.2} \times 10^{22}~{\rm cm}^{-2} \left({{0.3~{\rm keV}} \over {k_{\rm B}T_e}}\right)$ (95\% c.l.). This electron density is merely approximate due to the assumptions made in deriving $y$, and in presuming all of the measured signal is actually rooted in the SZ.

The column density of the free electrons is consistent with that suggested in Peiris \& Smith (2010)~\cite{Peiris:2010jd} to resolve the 
CMB isotropy anomalies via the kinetic SZ effect. In reality, both the thermal and kinetic SZ effects may contribute to large-scale anomalies 
of the CMB. However, the required column density is in strong tension with pulsar dispersion measurements (Taylor \& Cordes 1993~\cite{Taylor:1993my}), OVII absorption studies~\cite{Fang:2010yk}, and the soft X-ray background~\cite{Yoshino:2009kv}, which place $N_e \sim {10^{20}-10^{21}} \rm{cm}^{-2}$. 
Given the WMAP subtracted foregrounds would need to have been underestimated by a factor of two to explain our low-$\ell$ signal, this discrepancy suggests that the excess signal may originate from further instrumental effects or an unknown foreground. Rather than to simply explain away the signal, as we did not find either a simple statistical argument or systematic effect, we suggest that further studies are warranted.

\section{Acknowledgements}
We thank David Buote, Olivier Dor\'e, Taotao Fang, Gary Hinshaw, Fill Humphrey, Eiichiro Komatsu, Gregory Martinez, and Paolo Serra for helpful conversations. This work was supported by NSF AST0645427 and NASA NNX10AD42G.

\bibliographystyle{jhep}
\bibliography{text}

\end{document}